\documentclass[aps,prl,amsmath,amssymb,twocolumn,floatfix]{revtex4-1}
\usepackage{graphicx}
\usepackage{dcolumn}
\usepackage{bm}
\usepackage{xcolor}

\begin{document}

\title{Two-Gap Time Reversal Symmetry Breaking Superconductivity in Non-Centrosymmetric LaNiC$_2$}

\author{Shyam Sundar,$^1$ S. R. Dunsiger,$^{1,2}$ S. Gheidi,$^1$ K. S. Akella,$^1$ A. M.~C\^{o}t\'{e},$^1$ H. U.~\"{O}zdemir,$^1$ N. R. Lee-Hone,$^1$ 
D. M. Broun,$^1$ E. Mun,$^1$ F. Honda,$^3$ Y. J. Sato,$^3$ T. Koizumi,$^3$ R. Settai,$^4$ Y. Hirose,$^4$ I. Bonalde,$^5$ and J. E.~Sonier$^1$}

\affiliation{$^1$Department of Physics, Simon Fraser University, Burnaby, British Columbia V5A 1S6, Canada}
\affiliation{$^2$Centre for Molecular and Materials Science, TRIUMF, Vancouver, British Columbia V6T 2A3, Canada}
\affiliation{$^3$Institute for Materials Research, Tohoku University, Oarai, Ibaraki 311-1313, Japan}
\affiliation{$^4$Department of Physics, Niigata University, Niigata 950-2181, Japan}
\affiliation{$^5$Centro de F{\'i}sica, Instituto Venezolano de Investigaciones Cient{\'i}ficas, Apartado 20632, Caracas 1020-A, Venezuela}

\date{\today}
\begin{abstract}
We report a $\mu$SR investigation of a non-centrosymmetric superconductor (LaNiC$_2$) in single crystal form. Compared to
previous $\mu$SR studies of non-centrosymmetric superconducting polycrystalline and powder samples, the unambiguous orientation of single crystals 
enables a simultaneous determination of the absolute value of the magnetic penetration depth and the vortex core 
size from measurements that probe the magnetic field distribution in the vortex state. The magnetic field dependence of these quantities 
unambiguously demonstrates the presence of two nodeless superconducting energy gaps. In addition, we detect weak internal magnetic fields in the superconducting 
phase, confirming earlier $\mu$SR evidence for a time-reversal symmetry breaking superconducting state.
Our results suggest that Cooper pairing in LaNiC$_2$ is characterized by the same interorbital equal-spin pairing model introduced
to describe the pairing state in the centrosymmetric superconductor LaNiGa$_2$. 
\end{abstract}

\maketitle
While the formation of Cooper pairs in a superconductor is generally protected by time-reversal and inversion symmetries,
superconductivity is also exhibited by certain materials lacking a center of inversion in their crystal structure.
These so-called non-centrosymmetric superconductors (NCSCs) have garnered a great deal of attention in the past decade.
The lack of inversion symmetry enables an antisymmetric spin-orbit coupling (ASOC) of the single electron states, which
facilitates mixing of spin-singlet and spin-triplet configurations in the superconducting (SC) pair wavefunction \cite{Bauer:2012,Smidman:2017}.
The degree of mixing is dependent on the strength of the ASOC.

There is evidence from zero-field (ZF) $\mu$SR measurements that the SC state of some NCSCs break time-reversal symmetry (TRS) 
\cite{Hillier:2009,Biswas:2013,RPSingh:2014,Barker:2015,DSingh:2017,Shang:2018a,DSingh:2018a,DSingh:2018b,Shang:2018b,Shang:2020}. 
Such appears to be the case for the non-centrosymmetric ternary carbide compound LaNiC$_2$, where
measurements on a polycrystalline sample show a weak increase in the ZF-$\mu$SR relaxation rate at the SC
transition temperature ($T_c$) indicative of the formation of spontaneous magnetic fields \cite{Hillier:2009}.
The only TRS breaking states permitted by the low point group symmetry ($C_{2v}$) 
of the orthorhombic crystal structure of LaNiC$_2$ are those with non-unitary spin-triplet pairing \cite{Quintanilla:2010,Mukherjee:2014}.
These allowed pairing states are incompatible with strong ASOC and appreciable singlet-triplet mixing, and
have nodes in the associated SC energy gap function \cite{Quintanilla:2010,Mukherjee:2014}. However, different experiments 
have reached very different conclusions regarding the energy gap structure in LaNiC$_2$. While initial specific heat measurements
suggested an unconventional SC gap with point nodes \cite{Lee:1996}, a conventional isotropic BCS $s$-wave gap is supported by subsequent 
specific heat \cite{Pecharsky:1998} and nuclear quadrupole resonance \cite{Iwamoto:1998} studies, as well as a strong suppression of $T_c$ by 
Ce substitution for La \cite{Katano:2017}. There is also evidence for the existence of two nodeless SC gaps in LaNiC$_2$ from specific heat measurements and changes 
in the magnetic penetration depth with temperature, $\Delta \lambda (T)$, measured in the Meissner state via a tunnel diode oscillator technique \cite{Chen:2013}.
Still other measurements of $\Delta \lambda(T)$ by the same method support the earlier claim of point nodes in the SC energy 
gap \cite{Bonalde:2011,Landaeta:2017}.

To explain experiments on LaNiC$_2$ and the centrosymmetric superconductor LaNiGa$_2$ that indicate TRS breaking and fully-gapped behavior,
a novel non-unitary triplet SC pairing state has been proposed in which pairing occurs between electrons of the same spin, but on two different orbitals \cite{Weng:2016}. 
Depending on the character of the two orbitals involved in this interorbital equal-spin pairing (ESP) state, two nodeless SC energy gaps associated with the
two different values of the Cooper pair spin ($S_z \! = \! +1$ and $S_z \! = \! -1$) may exist. If the ESP occurs between electrons
on different orbitals of the Ni atom, a gap with line nodes or a single anisotropic SC energy gap may occur \cite{Csire:2018}. 
Only the latter requires the non-centrosymmetric crystal structure of LaNiC$_2$.

In this Letter, we resolve the question of the SC gap structure in LaNiC$_2$ via $\mu$SR measurements of single crystals
in the vortex state. Two-gap superconductivity is unambiguously identified in the magnetic field dependence of the fitted values of the 
absolute value of the magnetic penetration depth ($\lambda$) and the vortex core size. To date all evidence of TRS breaking in NCSCs by ZF-$\mu$SR
has come from experiments on powder or polycrystalline samples via the observation of an increase in the relaxation rate at $T_c$
that is small compared to that detected in single crystals of TRS breaking centrosymmetric superconductors UPt$_3$ \cite{Luke:1993}, Sr$_2$RuO$_4$ \cite{Luke:1998}, 
and PrOs$_4$Sb$_{12}$ \cite{Aoki:2003}. Here we also report evidence for TRS breaking in LaNiC$_2$ single crystals by the detection of
weak internal magnetic fields below $T_c$.

Single crystals of LaNiC$_2$ were grown by the Czochralski method, as described in Ref.~\cite{Hirose:2012}. 
Heat capacity and magnetization measurements indicate that bulk superconductivity occurs with a ZF value $T_c \! \sim \! 2.7$~K and an 
upper critical magnetic field $H_{c2}^{\parallel a} \! \sim \! 1.53$~kOe \cite{Supplemental}.
A secondary phase of La$_2$Ni$_5$C$_3$ identified by X-ray diffraction is present in $\sim \! 5$~\% of the sample volume,
but is non-superconducting down to at least 0.11~K \cite{Supplemental}.    
 
\begin{figure}
\includegraphics[scale=0.5]{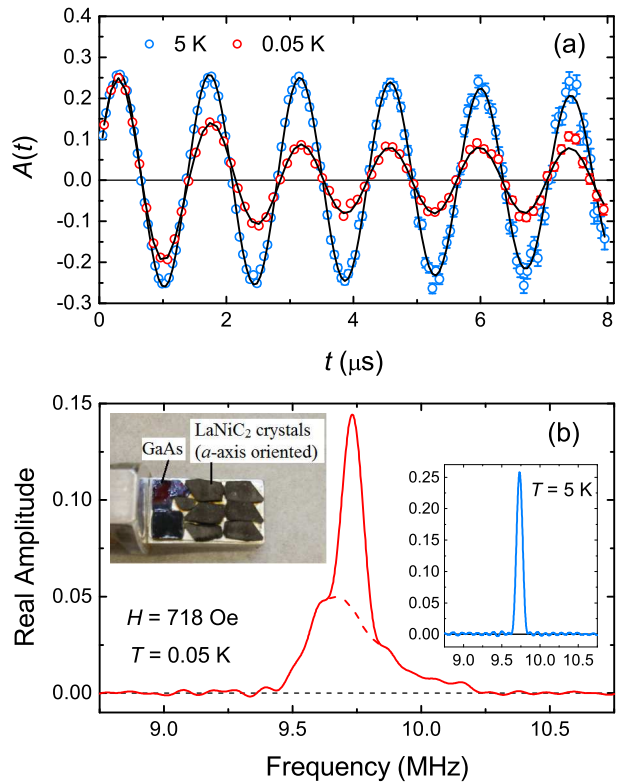}
\caption{(a) TF-$\mu$SR asymmetry spectra recorded above and below $T_c$ for a magnetic field of $H \! = \! 718$~Oe displayed
in a 9.03~MHz rotating reference frame. Note, the precession frequency is related to the local field by $\nu \! = \! (\gamma_\mu/2 \pi)B$,
where $\gamma_\mu/2 \pi \! = \! 13.5539$~MHz/kG is the muon gyromagnetic ratio.
Also shown are fits that use Eq.~(\ref{eq:1}) below $T_c$ \cite{Supplemental}. (b) Fourier transform of the TF-$\mu$SR signal 
for $T \! = \! 0.05$~K.  
The large peak at 9.73~MHz is due to muons stopping outside the sample. Left inset: Photograph of the LaNiC$_2$ single crystals 
and GaAs wafers attached to an Ag backing plate on an Ag sample holder. Right inset: Fourier transform of TF-$\mu$SR signal for $T \! = \! 5$~K.}
\label{fig1}
\end{figure}

The $\mu$SR experiments were performed on the M15 surface muon beamline at TRIUMF using a top loading dilution refrigerator. A mosaic of seven $a$-axis aligned LaNiC$_2$ 
single crystals, each $\sim \! 0.5$~mm thick and six having a mass of $\sim \! 70$~mg each,
were mounted on a 12.5~mm~$\times$~22~mm~$\times$~0.25~mm pure Ag plate and attached to an Ag sample holder. Together, the LaNiC$_2$ single crystals covered 
$\sim \! 70$~\% of the Ag plate (see Fig.~\ref{fig1}(b) inset).
Measurements in the vortex state were performed in a transverse-field (TF) geometry \cite{Sonier:2000}, with the magnetic field applied parallel to 
the $a$-axis of the LaNiC$_2$ single crystals and transverse to the initial muon-spin polarization {\bf P}$(t \! = \! 0)$. 
The field was first applied above $T_c$ before cooling into the vortex state. To reduce the contribution to 
the TF-$\mu$SR signal from muons stopping in the Ag backing plate, three thin wafers of intrinsic GaAs were used to cover the exposed area around the 
LaNiC$_2$ sample --- GaAs produces no detectable muon precession signal for the field range considered in our study. No GaAs was used for the
ZF measurements, as this would give rise to a relaxing signal associated with muonium (Mu $\equiv \! \mu^+$e$^-$) formation. Conversely, 
the very small Ag nuclear moments produce no appreciable relaxation of the ZF-$\mu$SR signal.
For the ZF-$\mu$SR experiments, {\bf P}$(t \! = \! 0)$ was parallel to the $a$-axis and stray external magnetic fields at the sample position 
were reduced to $\lesssim \! 35$~mG using field compensation coils and the precession signal of Mu in intrinsic Si as a sensitive magnetometer \cite{Morris:03}. 

Figure~\ref{fig1}(a) shows representative TF-$\mu$SR asymmetry spectra $A(t)$ displayed in a rotating reference frame. The significant damping of the signal 
below $T_c$ is due to muons randomly sampling the spatial distribution of magnetic field associated with a vortex lattice (VL).
Gaussian apodized Fourier transforms (FTs) of the TF-$\mu$SR signals are shown in Fig.~\ref{fig1}(b) and the Supplementary Material \cite{Supplemental}. 
The FT is representative of the magnetic field distribution $n(B)$ sensed by muons stopping inside and outside the sample, but is broadened by the apodization used to smooth out the 
ringing and noise artifacts caused by the finite time range and the reduced number of muon decay events at long times \cite{Sonier:2000}. The FT below $T_c$ shows a large peak due to muons stopping in the
Ag backing plate or sample holder, superimposed on an asymmetric lineshape generated by muons sensing the nuclear moments and VL in the 
LaNiC$_2$ single crystals. Below $T_c$, the VL contribution to $A(t)$ \cite{Supplemental} is well described by
the following analytical Ginzburg-Landau (GL) model for the spatial variation of field generated by a hexagonal VL \cite{Yaouanc:1997}       
\begin{equation}
B({\bf r}) = B_0 \left(1-b^{4}\right) \sum_{\bf G} \frac{e^{-i {\bf G} \cdot {\bf r}} u K_{1}(u)}{\lambda_{bc}^{2} G^{2}} \, ,
\label{eq:1}
\end{equation}
where $b \! = \! B/B_{c2}$ is the ratio of the local and upper critical magnetic fields, $B_0$ is the average internal magnetic field, {\bf G} are the VL reciprocal 
lattice vectors, $K_1(u)$ is a modified Bessel function, $u^2 \! = \! 2 \xi_{bc}^2 G^2 (1 + b^4)[1-2b(1 - b)^2]$, and $\xi_{bc}$ and $\lambda_{bc}$ are the coherence 
length and magnetic penetration depth associated with supercurrents flowing in the $bc$-plane. The suitability of Eq.~(\ref{eq:1}) has been widely demonstrated in
previous $\mu$SR investigations of type-II superconductors, where the vortex core size ($r_0$) is defined as the radial distance from the vortex center to the maximum in the 
absolute value of the supercurrent density $j(r) \! = \! | {\bf \nabla} \times {\bf B}({\bf r})|$ \cite{Sonier:2004,Sonier:2007}. 
Since changes in the slope of the pair potential $\Delta(r)$ in the vortex core region modify the cutoff factor $u K_1(u)$ in Eq.~(\ref{eq:1}), 
changes in the core size modify the fitted value of $\xi_{bc}$ \cite{Sonier:2007}. Consequently, the ``true'' GL coherence length is 
the value of $\xi_{bc}$ in the $T \! \rightarrow \! 0$ and $H \! \rightarrow \! 0$ limits.
 
\begin{figure}
\includegraphics[scale=0.55]{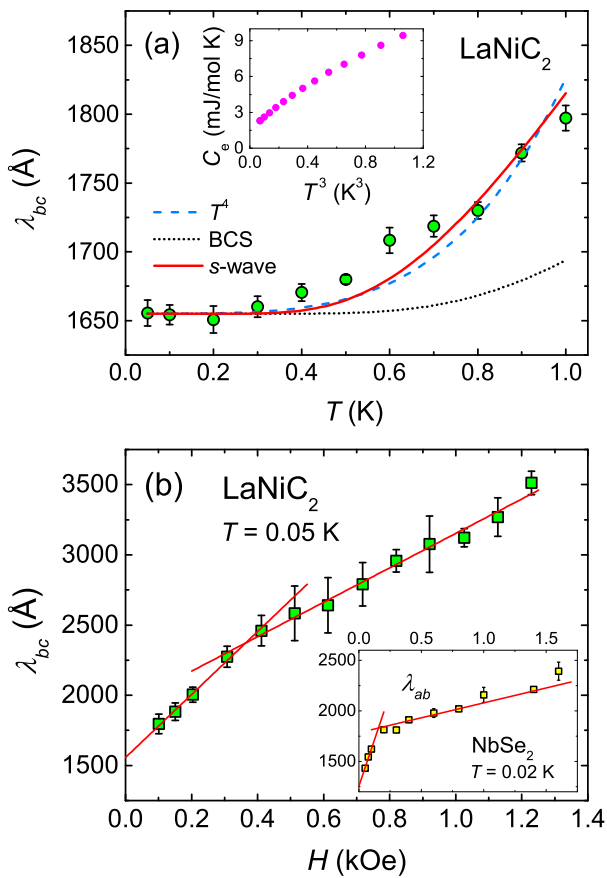}
\caption{(a) Temperature dependence of $\lambda_{bc}$ in LaNiC$_2$ below $T \! \sim \! 0.38 T_c$ and for $H \! = \! 150$~Oe. The dashed curve is a fit of the form $\lambda_{bc}(0) \! + \! a T^4$
expected for point nodes along the $a$ axis. The dotted and solid curves are fits to the single-gap $s$-wave BCS expression \cite{Halbritter:1971} with zero-temperature energy gap 
values $\Delta_{bc}(0) \! = \! 1.76 k_B T_c$ and $\Delta_{bc}(0) \! = \! 1.16(3) k_B T_c$, respectively. Inset: Low-temperature (0.41~K~$\leq \! T \! \leq \! 1$~K)
electronic contribution to the heat capacity plotted versus $T^3$. 
(b) Magnetic field dependence of $\lambda_{bc}$ in LaNiC$_2$ for $T \! = \! 0.05$~K. The straight lines are linear fits for $H \! \leq \! 0.4$~kOe and $H \! \geq \! 0.5$~kOe.
The fit for $H \! \leq \! 0.4$~kOe yields the ZF value $\lambda_{bc}(0) \! = \! 1548 \! \pm \! 24$~\AA. Inset: Magnetic field dependence of $\lambda_{ab}$ in NbSe$_2$
for $T \! = \! 0.02$~K \cite{Callaghan:2005}.}  
\label{fig2}
\end{figure}

Figure~\ref{fig2}(a) shows the temperature dependence of $\lambda_{bc}$ obtained from fits of TF-$\mu$SR spectra recorded for $H \! = \! 150$~Oe and $T \! \leq \! 1$~K. 
The data are poorly described by the form $\lambda_{bc}(T) \! - \! \lambda_{bc}(0) \! \propto \! T^4$ expected for point nodes along the 
$a$ axis \cite{Landaeta:2017}. Moreover, an accompanying $T^3$ dependence of the electronic specific heat could not be confirmed from measurements above 0.41~K (Fig.~\ref{fig2}(a) inset).
When fitting $\lambda_{bc}(T)$ to a single-gap $s$-wave BCS model \cite{Halbritter:1971}, with $\Delta_{bc}$ as an adjustable parameter,  
we infer a much smaller gap than the BCS value of $\Delta_{bc}(0) \! = \! 1.76 k_B T_c$. The small SC gap value may correspond to the minimum of a single anisotropic gap 
or the smallest gap of a multi-gap state. These results highlight the challenges when attempting to draw conclusions about the SC gap structure 
from fits of the temperature dependence of the magnetic penetration depth or thermodynamic quantities. 

By contrast, a clear indication of the gap structure in LaNiC$_2$ is provided by the low-$T$ magnetic field dependence
of $\lambda_{bc}$, which is displayed in Fig.~\ref{fig2}(b). The linear growth of $\lambda_{bc}$ with increasing $H$ and the change in slope 
above $H \! \sim \! 0.4$~kOe resembles the behaviour of $\lambda_{ab}(H)$ in the two-gap $s$-wave superconductor NbSe$_2$ \cite{Callaghan:2005}.
We note that $\lambda$ exhibits no field dependence in a single fully-gapped $s$-wave superconductor below 
$H/H_{c2} \! \sim \! 0.5$ \cite{Sonier:2004b,Kadono:2004} and displays a sublinear dependence on field in a superconductor with gap nodes \cite{Sonier:1999}.
In NbSe$_2$, the initial steep linear-$H$ increase of $\lambda_{ab}$ is attributed to the delocalization of loosely bound quasiparticle (QP) 
vortex core states associated with a small full energy gap on one of the conduction-electron bands. The delocalization results from an increased overlap of the
bound QP states of neighboring vortices, which occurs due to the increase in vortex density at higher field. This modifies $B(\bf{r})$ around the vortex 
cores and the fitted value of $\lambda_{ab}$ (or $\lambda_{bc}$). The ``true'' magnetic penetration depth for LaNiC$_2$ is the extrapolated 
value $\lambda_{bc}(T \! \rightarrow \! 0, H \! \rightarrow \! 0)$, determined to be $1548 \! \pm \! 24$~\AA~ from the linear fit of the low-field 
data presented in Fig.~\ref{fig2}(b). 

\begin{figure}
\includegraphics[scale=0.35]{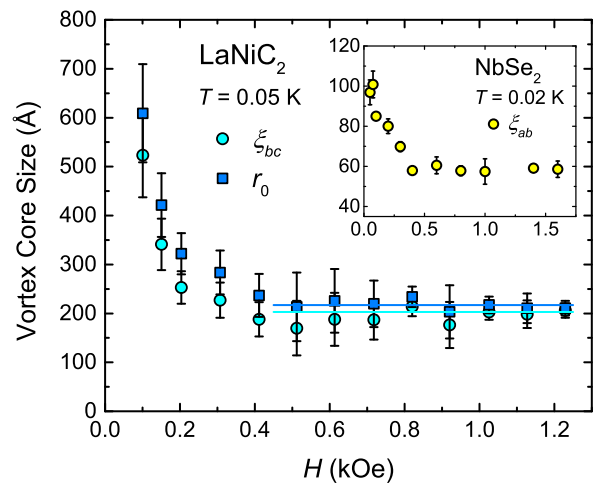}
\caption{Magnetic field dependence of $r_0$ and $\xi_{bc}$ in LaNiC$_2$ for $T \! = \! 0.05$~K. Inset: Magnetic field dependence of $\xi_{ab}$ in NbSe$_2$
for $T \! = \! 0.02$~K \cite{Callaghan:2005}.}
\label{fig3}
\end{figure}

The delocalization of QP core states in NbSe$_2$ at low $T$ leads to a rapid decrease in $r_0$ (and $\xi_{ab}$) with increasing $H$, before 
saturating at higher fields where the vortex structure is controlled by the larger full SC gap on a different conduction band \cite{Callaghan:2005}.
This is accompanied by a reduction in the slope of the linear-$H$ dependence of $\lambda_{ab}$.
The saturation of the core size is due to the QP core states being more tightly bound to the smaller vortices associated with the larger gap. 
As shown in Fig.~\ref{fig3}, the field dependence of the vortex core size in LaNiC$_2$ at low $T$ exhibits a behavior similar to that of NbSe$_2$,
which is distinct from the behavior of the core size in a single-gap $s$-wave superconductor \cite{Sonier:2004b}. 
As expected, the low-field value of $\xi_{bc}$ above the lower critical field ($H_{c1} \! \sim \! 0.1$~kOe) 
is close to the calculated value $\xi_{bc} \! = \! [\Phi_0/2 \pi H_{c2}^{\parallel a}]^{1/2} \! \approx \! 464$~\AA.

\begin{figure}
\includegraphics[scale=0.9]{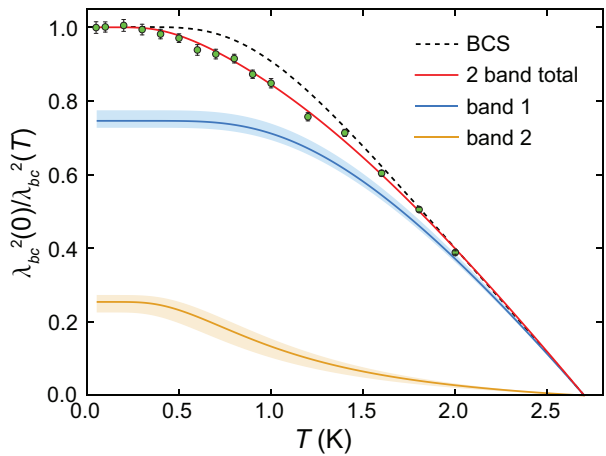}
\caption{Temperature dependence of the normalized superfluid density, $\lambda^2_{bc}(0)/\lambda^2_{bc}(T)$, in LaNiC$_2$ for $H = 150$~Oe. 
Circles denote $\mu$SR data points and error bars give the standard error at each temperature. The dashed curve is the superfluid density from 
single-band BCS theory. The upper solid curve is the total superfluid density in the two-band model, with contributions from the individual bands 
shown below it. Shaded areas denote the 1 $\sigma$ uncertainty regions associated with the model fit. Fit parameters are given in Ref.~\cite{Supplemental}.}
\label{fig4}
\end{figure}

To explore the existence of the two distinct SC gaps in more detail, we have fit the temperature dependence of the normalized low-field superfluid density, 
$\lambda^2_{bc}(0)/\lambda^2_{bc}(T)$, to a two-band, weak-coupling BCS model described in the Supplemental Material \cite{Supplemental}. Techniques that measure the temperature dependence of \emph{absolute} 
superfluid density, such as $\mu$SR, directly probe the thermal excitation of quasiparticles across the energy gaps, with the two-band fit revealing the underlying energy gaps 
in a tightly constrained manner. As shown in Fig.~\ref{fig4}, the single-band BCS curve does not adequately capture the measured superfluid density. The two-band model, however, 
provides a very good fit, and allows the contributions from the individual bands to be resolved. The range of temperature over which the
thermally activated behavior of each band appears to be temperature independent is indicative of the energy gap in each band. 
From the detailed temperature dependence of the energy gaps \cite{Supplemental},  
we infer zero-temperature gap ratios, $\Delta_i(0)/k_B T_c$, of 1.82 and 0.77, respectively.  
We note that fits of the data in Fig.~\ref{fig4} to a two-superconductor model confirm the absence of a second superconducting phase 
in our LaNiC$_2$ sample \cite{Supplemental}.   

\begin{figure}
\includegraphics[scale=0.35]{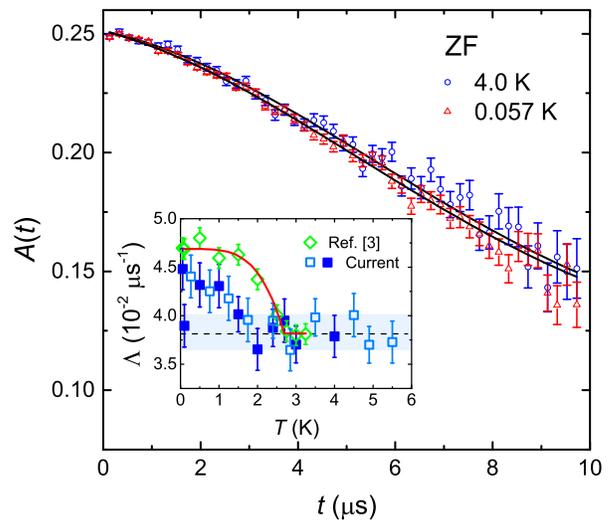}
\caption{Representative ZF-$\mu$SR asymmetry spectra. The solid curves are fits to Eq.~(\ref{eq:2}). 
Inset: Temperature dependence of the exponential ZF relaxation rate $\Lambda$. The open and solid squares correspond to two independent measurements of the
LaNiC$_2$ single crystals. The open diamonds are the exponential relaxation rate measured previously in polycrystalline LaNiC$_2$ \cite{Hillier:2009}.  
The dashed horizontal line denotes the average value of $\Lambda$ above $T_c$, $\langle \Lambda \rangle_{\rm N}$, for each data set. 
The data set from Ref.~\cite{Hillier:2009} and that denoted by the solid squares have been shifted vertically upward so that 
$\langle \Lambda \rangle_{\rm N}$ for all three data sets coincide.}
\label{fig5}
\end{figure}

Figure~\ref{fig5} shows ZF-$\mu$SR asymmetry spectra for our LaNiC$_2$ single crystals. These spectra are
reasonably described by the same function applied in the earlier ZF-$\mu$SR study of a polycrystalline sample \cite{Hillier:2009}
\begin{equation}
A(t) \! = \! A_0 G_{\rm KT}(\sigma, t) \exp(-\Lambda t) \! + \! A_{\rm bg} \, ,
\label{eq:2}
\end{equation} 
which consists of a relaxing term caused by the sample and a constant $A_{\rm bg}$ due to muons stopping in the Ag backing plate/sample holder. 
Here $G_{\rm KT}(\sigma, t)$ is a static Gaussian Kubo-Toyabe function \cite{Kubo:1967}. The ZF-$\mu$SR spectra were analyzed assuming only the relaxation rate
$\Lambda$ changes with temperature and with nearly equivalent values of $A_0$ and $A_{\rm bg}$ determined from weak TF-$\mu$SR measurements 
in the Meissner state.
The fits yield $\sigma \! = \! 0.104 \! \pm \! 0.002$~$\mu$s$^{-1}$ and the variation of $\Lambda$ with temperature 
displayed in the inset of Fig.~\ref{fig5}. Also shown is the increase of the ZF-$\mu$SR relaxation rate reported in Ref.~\cite{Hillier:2009}, 
which corresponds to a characteristic field strength of 0.10~G. While the error bars and scatter of our data are greater, an
increase of $\Lambda$ is observed at lower temperature.

The occurrence of spontaneous fields in LaNiC$_2$ is apparently sample dependent. In addition to the previous results for a 
polycrystalline sample \cite{Hillier:2009},
an extremely small $c$-axis aligned spontaneous magnetization ($\sim \! 10^{-5}$~G) far below the sensitivity of ZF-$\mu$SR 
has been observed to appear in a single crystal at $T_c$, but not in a second single crystal investigated in the same study \cite{Sumiyama:2015}.
The situation of LaNiC$_2$ somewhat resembles the lower TRS breaking SC phase of UPt$_3$, where 
weak spontaneous internal fields appearing at $T_c$ were first detected by ZF-$\mu$SR \cite{Luke:1993}, but not 
later in higher quality single crystals \cite{deReotier:1995,Higemoto:2000}.   
In TRS breaking superconductors, intrinsic spontaneous magnetism is generated near the surface and by impurities and defects that disturb the 
SC order parameter \cite{Sigrist:2005}. Our measurements and those of Ref.~\cite{Hillier:2009} are not sensitive to spontaneous currents at the surface, 
as the mean stopping depth of the muons in LaNiC$_2$ ($\sim \! 0.2$~mm) used in these experiments far exceeds the magnetic penetration depth.
We attribute the onset of weak internal fields at a lower temperature in our single crystals to the width of the SC transition.
While magnetization measurements show a diamagnetic response beginning at 2.7~K, the zero-field cooled curve does not saturate until $\sim \! 1.8$~K \cite{Supplemental}.
Given the small increase of the ZF-$\mu$SR relaxation rate that is observed, it is likely that our experiments are not sensitive to the intrinsic spontaneous magnetism 
until sufficent spontaneous currents are formed around the sample inhomogeneities. We note that spontaneous internal fields are observed at a temperature 
substantially lower than $T_c$ in ZF-$\mu$SR studies of other TRS breaking superconductors with broad SC transitions \cite{Bhattacharyya:2015,Singh:2018}

In summary, we have demonstrated the existence of two full SC gaps in LaNiC$_2$ via detection of the field dependence
of the magnetic field distribution at low temperatures in the vortex state of single crystals. Combined with supporting evidence for 
TRS breaking in the same sample, the nodeless two-gap SC state in LaNiC$_2$ is compatible with inter-orbital equal spin Cooper pairing \cite{Weng:2016}. 
         
 \acknowledgments{We thank Manfred Sigrist for insightful discussions and the staff of TRIUMF's Centre for Molecular and Materials Science for 
technical support. J.E.S., S.R.D., D.M.B. and E.M. acknowledge support from the Natural Sciences and Engineering Research Council (NSERC) of Canada. 
This research is also supported by the Japan Society for the Promotion of Science (JSPS) KAKENHI under grants JP15K05156 and JP15KK0149.}  

\bibliographystyle{apsrev}

\clearpage

\section*{Supplemental Material}



\date{\vspace{-5ex}}

\maketitle

\subsection{Powder X-ray diffraction spectrum of LaNiC$_2$}

\begin{figure}[h]
\centering
\includegraphics[scale=0.3]{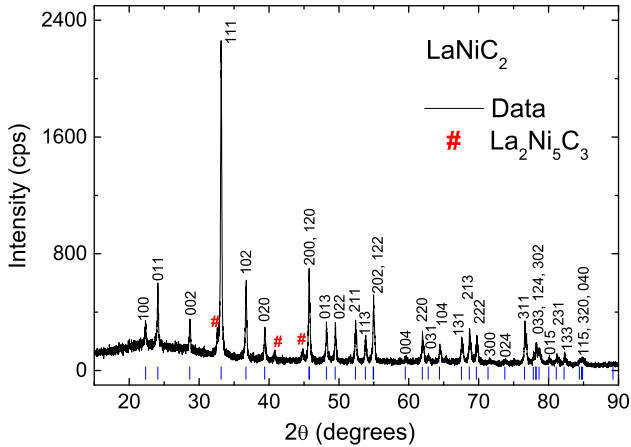}
\caption{Powder X-ray diffraction (XRD) spectrum of LaNiC$_2$. The blue vertical lines indicate the calculated Bragg peak 
positions for LaNiC$_2$, which crystallizes in the orthorhombic $Amm2$ space group. The three low intensity XRD Bragg peaks 
labelled with $\#$ are associated with an $\sim \! 5$~\% secondary phase of La$_2$Ni$_5$C$_3$, which crystallizes in the tetragonal $P4/mbm$ space group.}
\label{figS1}
\end{figure}

X-ray diffraction (XRD) measurements were performed on one of the single crystals of LaNiC$_2$ ground to a fine powder using a conventional X-ray 
diffractometer (RINT-2200, Rigaku Co. Ltd) equipped with a Cu target source. In addition to sharp Bragg peaks associated
with LaNiC$_2$, the XRD spectrum (Fig.~\ref{figS1}) shows three low intensity peaks associated with a La$_2$Ni$_5$C$_3$ 
secondary phase occupying $\sim \!5$~\% of the sample. The lattice parameters for LaNiC$_2$ determined from the XRD spectrum are $a \! = \! 3.9578$~\AA, 
$b \! = \! 4.5616$~\AA~ and $c \! = \! 6.2001$~\AA, which are in good agreement with literature values \cite{Lee1996, Chen2013}. 

\section{Powder X-ray diffraction spectrum and resitivity of La$_2$Ni$_5$C$_3$}

\begin{figure}[h]
    \centering
\includegraphics[scale=0.3]{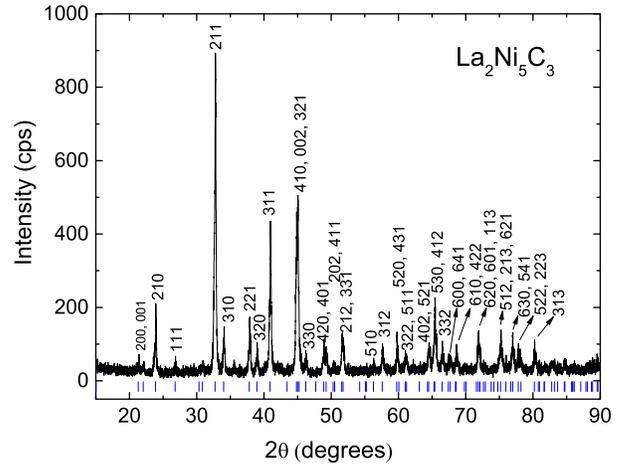}
\caption{Powder X-ray diffraction (XRD) spectrum of La$_2$Ni$_5$C$_3$. The blue vertical lines indicate 
the calculated Bragg peak positions for La$_2$Ni$_5$C$_3$, which crystallizes in the tetragonal $P4/mbm$ space group.
The lattice parameters for La$_2$Ni$_5$C$_3$ determined from the XRD spectrum are $a \! = \! b \! = \! 8.3266$~\AA~ and $c \! = \! 4.0244$~\AA.}
\label{figS2}
\end{figure}

\begin{figure}[h]
    \centering
\includegraphics[scale=0.3]{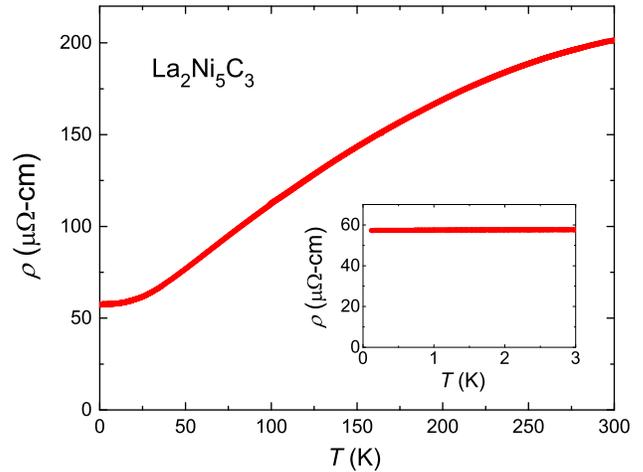}
\caption{Temperature dependence of the resistivity of polycrystalline La$_2$Ni$_5$C$_3$ down to 0.11~K.}
\label{figS3}
\end{figure}

Previous studies show that La$_2$Ni$_5$C$_3$ is non-magnetic and does not exhibit superconductivity down to 1.8~K \cite{Kato2006,Jeitschko1989}.
To determine whether superconductivity exists at lower temperatures, we synthesized a polycrystalline sample of La$_2$Ni$_5$C$_3$. 
A powder XRD spectrum of the La$_2$Ni$_5$C$_3$ sample is shown in Fig.~\ref{figS2}.
As shown in Fig.~\ref{figS3}, no superconducting transition is observed in the temperature dependence of the resistivity down to 0.11~K.

\newpage

\section{Heat capacity and magnetization measurements of LaNiC$_2$}

\begin{figure}[h]
\centering
\includegraphics[scale=0.32]{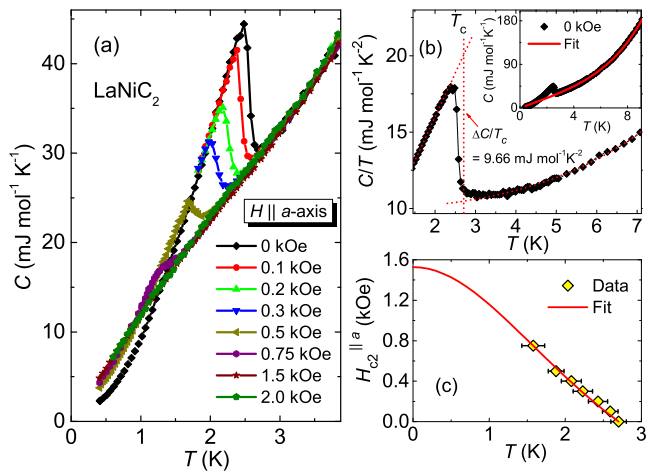}
\caption{(a) Temperature dependence of the heat capacity for different values of magnetic field applied parallel to the $a$-axis. 
(b) The dotted lines duplicate a procedure described in Ref.~\cite{Budko2009} to estimate $T_c$ and the heat capacity jump 
$\Delta C/ T_c$ at $T_c$. Inset shows a 
polynomial fit to the normal state heat capacity data for zero field over the temperature range 3.1~K to 10.2~K, as 
described in the main text.  
(c) Temperature dependence of the upper critical field $H_{c2}^{\parallel a}(T)$ estimated from the heat capacity data in (a). 
The solid red curve is a fit to the data, which is described in the main text.}
\label{figS4}
\end{figure}

Heat capacity measurements were performed on a small piece of one of the LaNiC$_2$ single crystals down to 0.4~K using 
a Quantum Design Physical Property Measurement System. The temperature dependence of the heat capacity $C(T)$ measured in 
zero and different applied magnetic fields is shown in Fig.~\ref{figS4}(a).
The bulk superconducting transition temperature $T_c$ and the heat capacity jump at $T_c$ are determined 
in Fig.~\ref{figS4}(b) by a procedure described in Ref.~\cite{Budko2009}, which yields $T_c \! \sim \! 2.7$~K and 
$\Delta C/ T_c \! \sim \! 9.66$ mJ mol$^{-1}$ K$^{-2}$. These are consistent with literature values \cite{Lee1996, Chen2013, Pecharsky1998}.
The dotted curve through the data above $T_c$ in Fig.~\ref{figS4}(b) comes from fitting the zero-field heat capacity 
from 3.1~K to 10.2~K to a polynomial function, $C(T) \! = \!  \gamma T \! + \! \beta T^3 \! + \! \delta T^5$, 
as shown in the inset of Fig.~\ref{figS4}(b). The first term 
describes the electronic contribution to the heat capacity and the last two terms describe the lattice contribution. 
The fit yields $\gamma \! \sim \! 9.9$~mJ mol$^{-1}$ K$^{-2}$ and $\beta \! \sim \! 0.066$~mJ mol$^{-1}$ K$^{-4}$.
A fit of the data from 6~K to 20~K instead yields $\gamma \! \sim \! 7.7$~mJ mol$^{-1}$ K$^{-2}$ and 
$\beta \! \sim \! 0.142$~mJ mol$^{-1}$ K$^{-4}$, which are in good agreement with previously reported values 
\cite{Lee1996, Chen2013, Pecharsky1998}. The heat capacity jump $\Delta C/ \gamma T_c \! \sim \! 0.98$ and 1.25 for 
$\gamma \! \sim \! 9.9$~mJ mol$^{-1}$ K$^{-2}$ and $7.7$~mJ mol$^{-1}$ K$^{-2}$, respectively.

Figure~\ref{figS4}(c) shows the temperature dependence of 
the upper critical field $H_{c2}^{\parallel a}$ estimated from the heat capacity data in Fig.~\ref{figS4}(a). 
The solid curve is a fit to the empirical relation $H_{c2}^{\parallel a}(T) \! = \! H_{c2}^{\parallel a}(0)[1-(T/T_c)^2]/[1+(T/T_c)^2]$,
yielding values $H_{c2}^{\parallel a}(0) \! = \! 1.53$~kOe and $T_c \! = \! 2.7$~K that are in good agreement with a
previous study of single crystal LaNiC$_2$ \cite{Hirose2012}.  


\begin{figure}[h]
\centering
\includegraphics[scale=0.32]{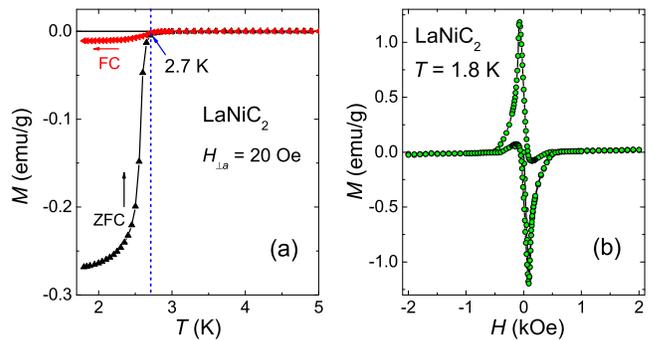}
\caption{(a) Temperature dependence of the dc magnetization down to 1.8 K measured for a 20~Oe magnetic field applied
perpendicular to the $a$-axis under zero-field cooled (ZFC) and field-cooled (FC) conditons.
(b) Magnetic hysteresis ($M$ versus $H$) loop at 1.8~K.}
\label{figS5}
\end{figure}

Measurements of the dc magnetization of LaNiC$_2$ were performed using a Quantum Design Magnetic Property Measurement System. 
Zero-field cooled (ZFC) data shows the onset of bulk superconductivity at $\sim \! 2.7$~K [Fig.~\ref{figS5}(a)]. 
Figure~\ref{figS5}(b) shows a magnetization hysteresis loop for LaNiC$_2$ at 1.8 K. The magnetization versus magnetic field is
characteristic of a type-II superconductor. The small hysteresis suggests any pinning of vortices by inhomogeneities in the 
sample is weak.

\newpage

\section{Analysis of the TF-$\mu$SR asymmetry spectrum}

\begin{figure}[h]
\centering
\includegraphics[scale=0.45]{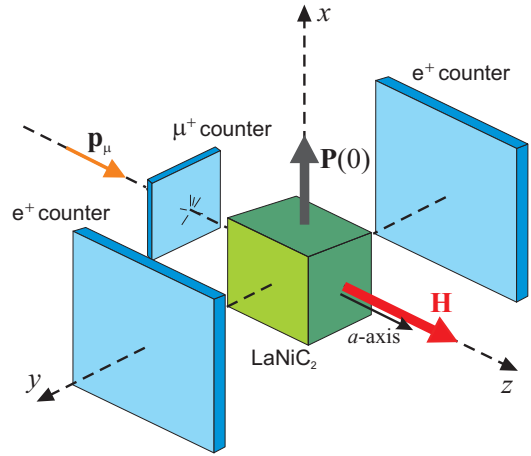}
\caption{Schematic of the geometry for the TF-$\mu$SR experiments. The time evolution of the muon spin polarization $P_x(t)$ is monitored via
detection of the muon decay positrons in a pair of counters positioned on opposite sides of the sample.}
\label{figS6}
\end{figure}

In the TF-$\mu$SR configuration used for our experiments (see Fig.~\ref{figS6}), the magnetic field {\bf H} was applied parallel to the muon beam momentum ${\bf p}_\mu$ ($z$ direction)
along the $a$-axis of the LaNiC$_2$ single crystals and the initial muon spin polarizaton ${\bf P}(t \! = \! 0)$ was rotated perpendicular to the applied field (in the $x$ direction). 
The TF-$\mu$SR asymmetry spectrum $A(t) \! = \! a_0 P_x(t)$ is the sum of sample and background contributions
\begin{equation}
A(t) = A_{\rm s}(t) + A_{\rm bg}(t) \,  .   
\end{equation}
The first term originates from muons that stop in the LaNiC$_2$ single crystals. The second term originates from muons 
that stop outside the sample in the Ag backing plate or sample holder. Muons stopping in the sample and those stopping in the Ag backing plate/sample holder sense 
a distribution of nuclear dipole fields that cause depolarization of the TF-$\mu$SR signal. Above $T_c$, the TF-$\mu$SR asymmetry spectrum 
is well described by the sum of two Gaussian damped cosine functions   
\begin{eqnarray}
A(t)  = & A_{\rm s}\exp(\sigma_{\rm s}^2 t^2) \cos(2 \pi \nu_{\rm s} t + \Phi) \nonumber \\
& + A_{\rm bg}\exp(\sigma_{\rm bg}^2 t^2)\cos(2 \pi \nu_{\rm bg} t + \Phi) \, .    
\end{eqnarray}
The precession frequencies $\nu_i$ ($i \! =$ s, bg) are a measure of the mean local field $B_i$ sensed by the muon, where $\nu_i \! = \! (\gamma_\mu/2 \pi)B_i$
and $\gamma_\mu/2 \pi \! = \! 13.5539$~MHz/kG is the muon gyromagnetic ratio. The parameter $\Phi$ is the initial phase of the muon spin polarization relative
to the positron counter axis ($y$-axis), which depends on the degree of Larmor precession of the muon spin in the applied field before reaching the sample.

Below $T_c$, the muons stopping in LaNiC$_2$ also sense the spatial variation in magnetic field caused by a vortex lattice and
the corresponding TF-$\mu$SR asymmetry spectra were fit to the following two-component depolarization function
\begin{eqnarray}
A(t) & = & A_{\rm s}\exp(\sigma_{\rm s}^2 t^2) \int_0^\infty n(B) \cos(\gamma_\mu B t + \Phi) dB \nonumber \\ 
& + & A_{\rm bg}\exp(\sigma_{\rm bg}^2 t^2)\cos(2 \pi \nu_{\rm bg} t + \Phi) \, ,    
\end{eqnarray}
where $n(B^\prime) \! = \! \langle \delta [B^\prime - B({\bf r})] \rangle$ is the probability of a muon sensing a local magnetic field $B$ in the $z$ direction
(parallel to the $a$-axis of the the LaNiC$_2$ single crystals) at a position {\bf r} in the $xy$-plane ($bc$-plane). The spatial field profile associated with the
vortex lattice $B({\bf r})$ was assumed to be described by Eq.~(1) in the main paper. Below $T_c$, the depolarization function $\exp(\sigma_{\rm s}^2 t^2)$
accounts for the effects of both the nuclear dipole fields and vortex lattice disorder on the internal magnetic field distribution. 

\newpage

\section{Representative Fourier transforms of TF-$\mu$SR asymmetry spectra}

\begin{figure}[h]
\centering
\includegraphics[scale=0.4]{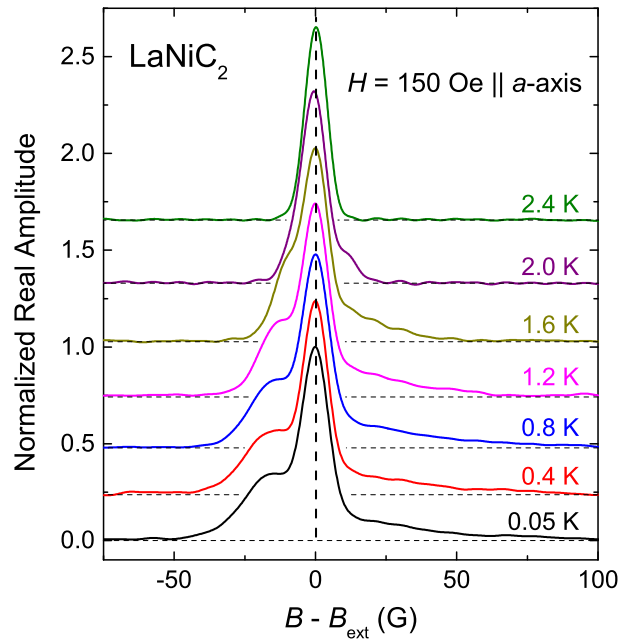}
\caption{Fourier transforms of the TF-$\mu$SR asymmetry spectrum in LaNiC$_2$ single crystals for a magnetic field of $H \! = \! 150$~Oe applied parallel to the $a$-axis.
The horizontal axis is the difference between the local field sensed by the muon and the external field.
The peak at $B \! - \! B_{\rm ext} \! = \! 0$ is a background signal originating from muons stopping outside the sample.}
\label{figS7}
\end{figure}

\newpage

\begin{figure}[h]
\centering
\includegraphics[scale=0.4]{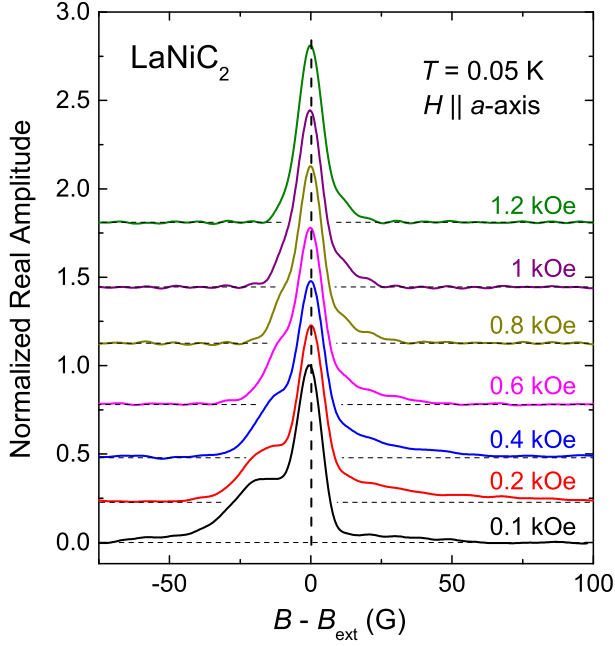}
\caption{Fourier transforms of the TF-$\mu$SR asymmetry spectrum in LaNiC$_2$ single crystals for $T \! = \! 0.05$~K and different values of the magnetic field applied 
parallel to the $a$-axis.}
\label{figS8}
\end{figure}

\section{Two-band BCS theory}

\subsection{Gap equation}

\begin{figure}[h]
    \centering
    \includegraphics[width = 3 in]{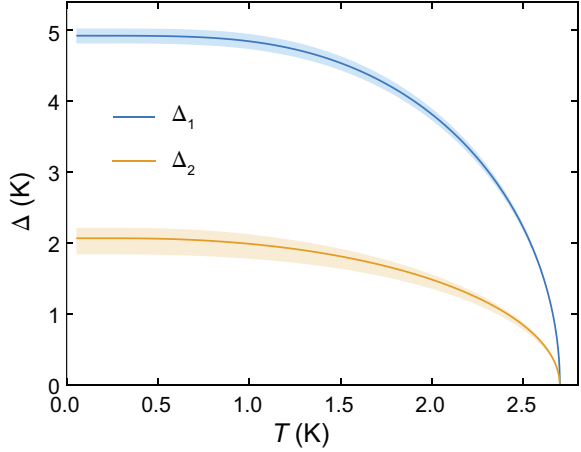}
    \caption{Temperature dependence of the energy gaps, $\Delta_1$ and $\Delta_2$, in the two-band model. Shaded areas denote the 1 sigma 
             uncertainty regions associated with the model fit.  The zero-temperature gap ratios, $\Delta_i(0)/k_B T_c$, are 1.82 and 0.77, respectively.}
    \label{figS9}
\end{figure}

In the Matsubara formalism, the temperature dependent gap equation for a weak-coupling superconductor is \cite{KMP}

\begin{equation}
\Delta_{\mathbf{k}}=2 \pi T N_0 \sum_{\omega_{n}>0}^{\omega_{0}}\left\langle V_{\mathbf{k}, \mathbf{k}^{\prime}} \frac{\Delta_{\mathbf{k}^{\prime}}}{\sqrt{\Delta_{\mathbf{k}^\prime}^{2} + \hbar^{2} \omega_{n}^{2}}}\right\rangle_{\!\!\mathrm{FS}}
\end{equation}
where $\omega_n = 2 \pi T (n + \frac{1}{2})$ are the fermionic Matsubara frequencies, $\Delta_\mathbf{k}$ is the gap parameter at wave vector $\mathbf{k}$, $N_0$ is the two-spin density of states, $V_{\mathbf{k}, \mathbf{k}^{\prime}}$ is the pairing interaction, $\langle...\rangle_\mathrm{FS}$ denotes an average over the Fermi surface and $\omega_0$ is a high frequency cutoff.  

The two-band superconductor describes situations in which the gap variation over the Fermi surface is approximately bimodal and can be approximated by two distinct gap scales, $\Delta_1$ and $\Delta_2$, one for each band. As discussed in Ref.~\cite{KMP}, the Fermi surface average is replaced by a sum over bands, and the pairing interaction is parameterized by a $2 \times 2$ symmetric matrix $\lambda_{\mu \nu}$, with the diagonal terms $\lambda_{11}$ and $\lambda_{22}$ describing intraband pairing, and the off-diagonal terms $\lambda_{12} = \lambda_{21}$ the interband interaction.  The gap equation then takes the simplified form
\begin{equation}
\Delta_{\nu}=\sum_{\mu=1,2} n_{\mu} \lambda_{\nu \mu} 2 \pi T \sum_{\omega_{n}>0}^{\omega_0}\frac{ \Delta_{\mu}}{\sqrt{\Delta_{\mu}^{2} +\hbar^{2} \omega_{n}^{2}}}\;,
\label{eqn:gapTemperature}
\end{equation}
where the relative densities of states for each band, $n_\mu$, obey $n_1 + n_2 = 1$.
For a given choice of parameters $\{n_1,\lambda_{11},\lambda_{22},\lambda_{12}\}$, Eq.~(\ref{eqn:gapTemperature}) is solved numerically, from which we obtain the temperature dependence of the gap parameters $\Delta_1$ and $\Delta_2$, as shown, for example, in Fig.~\ref{figS9}.

\subsection{Superfluid density}

Superfluid density is a thermal equilibrium property of the superconductor and is readily obtained within the Matsubara formalism once the energy gaps are known.  For band $\nu$, the normalized superfluid density is 
\begin{equation}
\rho_{\nu}(T)= \frac{\lambda_\nu^2(0)}{\lambda_\nu^2(T)} = \sum_{\omega_{n}>0}\frac{\Delta_{\nu}^{2} }{\left(\Delta_{\nu}^{2} +\hbar^2\omega_n^{2}\right)^{3/2}}\;.
\label{eqn:SFDensity}
\end{equation}
The total normalized superfluid density is a weighted sum of the contributions from each band,
\begin{equation}
\rho(T)=\gamma \rho_{1}(T)+(1-\gamma) \rho_{2}(t)\;,
\label{eqn:totalSF}
\end{equation}
where the weighting factor $0 < \gamma < 1$ is determined by the plasma frequency imbalance between the bands. Note that $\gamma$ is distinct from the density of states parameter $n_1$, as it includes Fermi velocity information: 
\begin{equation}
\gamma=\frac{n_{1} v_{1}^{2}}{n_{1} v_{1}^{2}+n_{2} v_{2}^{2}}\;,
\end{equation}
where $v_1$ and $v_2$ are the rms Fermi velocities of the two bands.

\subsection{Fitting procedure and results}

A least-squares optimization is used to search for best-fit parameters in the four-dimensional parameter space $\{n_1,\lambda_{11},\lambda_{22},\lambda_{12} \}$.  
For each parameter choice, the band-specific energy gaps and superfluid densities are determined at each of the experimental temperatures via numerical solution of 
Eqs.~(\ref{eqn:gapTemperature}) and (\ref{eqn:SFDensity}). As shown in Eq.~(\ref{eqn:totalSF}), the total superfluid density is a weighted combination of the band-specific superfluid densities.
While the weighting coefficient $\gamma$ is formally an additional fit parameter, a closed-form expression exists for its optimal value, so that it need not be included in the minimization search.  
$\gamma_\mathrm{opt}$ is found by minimizing the $\chi^2$ merit function
\begin{equation}
\begin{aligned}
\chi^{2} &=\left|\frac{\vec{\rho}_{\text {expt }}-\vec{\rho}_{\text {model }}}{\vec{\sigma}}\right|^{2} \\
&=\left|\frac{\vec{\rho}_{\text {expt }}-\vec{\rho}_{2}-\gamma \Delta \vec{\rho}}{\vec{\sigma}}\right|^{2} \\
&=\frac{\gamma^{2}|\Delta \vec{\rho}|^{2}-2 \gamma \Delta \vec{\rho} \cdot\left(\vec{\rho}_{\text {expt }}-\vec{\rho}_{2}\right)+\left|\vec{\rho}_{\text {expt }}-\vec{\rho}_{2}\right|^{2}}{|\vec{\sigma}|^2}
\end{aligned}
\end{equation}
where $\Delta \vec{\rho}=\vec{\rho}_{1}-\vec{\rho}_{2}$. Here the vector quantities encode the discrete temperature dependences of the various quantities, including experimental and model 
superfluid densities, and the measurement errors $\vec \sigma$. Minimizing with respect to $\gamma$ we obtain
\begin{equation}
\gamma_{\mathrm{opt}}=\frac{\Delta \vec{\rho} \cdot\left(\vec{\rho}_{\mathrm{expt}}-\vec{\rho}_{2}\right)}{|\Delta \vec{\rho}|^{2}}.
\end{equation}
In practice, the optimization depends only weakly on the choice of density of states parameter $n_1$.  Motivated by band-structure calculations \cite{Subedi2009, Hase2009} we 
estimate $0.8 \lesssim n_1 \lesssim 0.9$ to be an appropriate physical choice, and present results in Table~1 for $n_1 = 0.8$ and $n_1 = 0.9$.  
Figure~4 in the main paper and Fig.~\ref{figS9} here show the fits and gaps for $n_1 = 0.8$, and are practically indistinguishable from those for $n_1 = 0.9$.  
Note that while the $\lambda_{22}$ parameter appears to vary sharply between the two cases, it is the combination $n_2 \lambda_{22}$ that determines the intraband pairing strength 
in the second band, and this combination remains approximately constant. From this we conclude that the intrinsic pairing strength in the subdominant band is over an order of 
magnitude weaker than in the dominant band, and that interband pairing is important to the overall superconductivity.

\begin{figure}[h]
    \centering
    \includegraphics[width = 2.8 in]{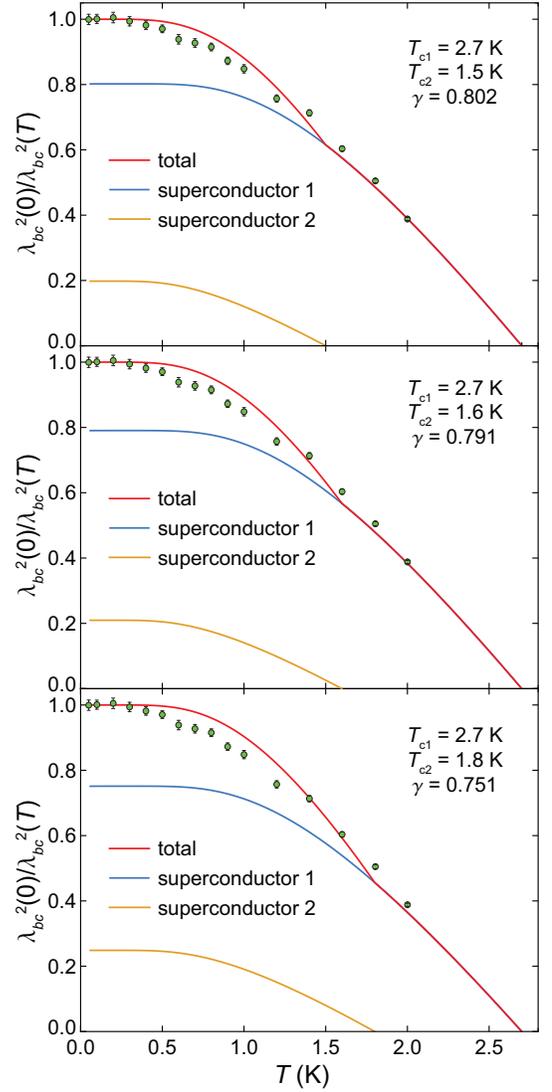}
    \caption{Fits to measured superfluid density in the two-phase superconductor model.  $T_{c1}$ is fixed at 2.7~K and $T_{c2}$ is varied, taking on representative values of 1.5~K, 1.6~K and 1.8~K in the three plots.}
    \label{figS10}
\end{figure}
 
\begin{table}[h!]
  \begin{center}

    \label{tab:table1}
    \begin{tabular}{|c||c|c||c|c|} 
    \hline
      fit parameter & $n_1 = 0.9$ & uncertainty &$n_1 = 0.8$ & uncertainty  \\
      \hline
      $\lambda_{11}$ & 0.77 & $\pm 0.012$ & 0.71 & $\pm 0.036$\\
      $\lambda_{22}$ & 0.26 & $\pm 0.013$ & 0.56 & $\pm 0.023$\\
      $\lambda_{12}$ & 0.23 & $\pm 0.018$ & 0.22 & $\pm 0.006$\\
      $\gamma_\mathrm{opt}$ & 0.747 & $\pm 0.029$ & 0.746 & $\pm 0.021$\\ \hline
    \end{tabular}
    \caption{Best-fit parameters and their uncertainties, for $n_1 = 0.8$ and $n_1 = 0.9$.}
  \end{center}
\end{table}

\subsection{Two-phase superconductor}

For comparison, we consider the superfluid density of a two-phase superconductor, which would apply if the material contained a secondary chemical phase that 
was also intrinsically superconducting.  In this scenario, the two superconducting phases have different superconducting transition temperatures, $T_{c1}$ and $T_{c2}$.  
Since there is no significant coupling between the phases, the temperature-dependent gap for each phase is the solution of a single-band BCS gap equation,
\begin{equation}
\Delta_{\nu}=  2 \pi T N_0 V_0 \sum_{\omega_{n}>0}^{\omega_0}\frac{ \Delta_{\nu}}{\sqrt{\Delta_{\nu}^{2} +\hbar^{2} \omega_{n}^{2}}}\;.
\label{eqn:singleBandGap}
\end{equation}
The corresponding superfluid density is still given by Eq.~(\ref{eqn:SFDensity}), and the effective superfluid density for the sample is still 
the weighted sum given by Eq.~(\ref{eqn:totalSF}). Fits to the measured absolute density are carried out by varying the $\gamma$ parameter, which controls 
the relative weighting of the two phases.  $T_{c1}$ is fixed at 2.7~K and $T_{c2}$ takes on representative values of 1.5~K, 1.6~K and 1.8~K.  
Results are shown in Fig.~\ref{figS10}. In each case, the best-fit superfluid densities in the two-phase scenario provide a much worse description 
of the data than the two-band model. Furthermore, the best fits require the secondary phase to contribute 20~\% to 25~\% of the total superfluid density, 
considerably greater than the 5~\% volume fraction of the known La$_2$Ni$_5$C$_3$ phase, which is in any case nonsuperconducting.

\end{document}